## Analysis of Bohr formula of momentum of inertia for even-even atomic nuclei

Mohd Kh. M. Abu El-Sheikh[1], Abdurahim Okhunov[2],

[1]*Depertment of Physics, Faculty of Science, University of Malaya, 50603 Kuala Lumpur, Malaysia*
[2]*Department of Science in Engineering, Kulliyyah of Engineering, International Islamic University Malaysia, 53100 Kuala Lumpur, Malaysia*
mhsr70@gmail.com, aaokhun@yahoo.com

***Key words:*** *Momentum of inertia, Hydrodynamical model, Irrotational motion*

## Abstract

The moment of inertia of even-even deformed nuclei which are derived on the basis of hydrodynamical model yield values that are too small compared with the experimental [Davidson 1965] ones. We expect that these contradictions come from the consideration only the first term in *R*– expansion in spite of not containing parameters indicating deformity of the nucleus, and neglecting all other terms which include the deformation parameters αλμ. In this work, the first three terms in *R*– expansion are taken into account. The results are more realistic and too much better the previous ones.

## A. Introduction

The problem of calculating the moment of inertia of even-even nuclei has received a considerable attention[1-3]. In the early discussion, the nucleus was regarded as a droplet of incompressible irrotational fluid and accordingly Bohr derived a simple formula $\mathcal{J} = 3B\beta^2$ used in the evaluation of the moment of inertia. This formula yield values nearly five times less than the empirical ones. This poor agreement lies in the concept of a spherical nucleus with small amplitude vibrations. Rayleigh[4] assumed these amplitudes are small to the extent that $R$ which is the upper limit of the integral in eq. (11) could be approximated to $R_0$, the radius of the nucleus, and hence he derived the equation $T = \frac{1}{2}\sum_{\mu} B_2|\dot{\alpha}_{2\mu}|^2$ for the kinetic energy, where $B_2 = \frac{\rho_0 R_0^5}{2}$ is the inertial parameter which was showed later it is unrealistic. Within the concept of irrotational flow, Villar[5] discussed this problem microscopically by taking the sum of the kinetic energy of all nucleons. He used in his study the canonical transformation and he obtained a mathematical formula for the moment of inertia that is similar to Bohr equation. Later, the kinetic energy was expressed in the form of power series in $(\alpha,\dot{\alpha})$ in tensor notation, we will quote it as it is [6]:

$$T(\pi,\alpha) = \frac{\sqrt{5}}{2B}\left[\pi^{[2]} \times \pi^{[2]}\right]^{[0]} + B_3[[\pi^2 \times \alpha^2]^2 \times \pi^2]^0 + \cdots$$

H. L. Acker and M. Marchal used the first two terms in this equation, they derived the formula $\mathcal{J}_k = 4\pi\beta^2\left[1 - 2B'\beta\cos\left(\gamma - k.\frac{2\pi}{3}\right)\right]\sin^2\left(\gamma - k.\frac{2\pi}{3}\right)$ for the moment of inertia, where $k = 1,2,3$ denotes to the body fixed coordinates.

If we assume a deformed nucleus with small vibrations about its equilibrium deformation many of these approximations that are taken into account by Rayleigh will be abandoned. The aim of this work is to discuss the effect of the second and third terms independently on the theoretical values of the moment of inertia However, in the remaining of this section, we will through a light on the main equations be used in our work. The work of Bohr will be outlined in section B, in section C and D the mathematical formulation will be discussed and finally, the calculation and discussion will be in section E. this part is found in more detail in books[7-10]

In the case of pure quadratic deformation $\lambda = 2$, the surface of the nucleus in a system of coordinate fixed in laboratoty is given by:

$$R = R_0\left(1 + \sum_\mu \alpha^*_{2\mu} Y_{2\mu}(\theta, \phi)\right). \quad (1)$$

where: $R$ is the radial coordinates of the surface in the direction $(\theta, \phi)$, $R_0$ is the radius of the surface when $\alpha^*_{\lambda\mu}$ is zero, $\mu = -2, \dots, 2$ describes the orientation of the motion with respect to space fixed coordinates and $\alpha^*_{\lambda\mu}$ is deformation coefficients (sometimes are used as collective coordinates).

$\alpha^*_{\lambda\mu}$ has the following properties

$$\alpha^*_{\lambda\mu} = (-1)^\mu \alpha_{\lambda,-\mu}, \text{ the conjugate property} \quad (2)$$

$$\alpha_{\lambda\mu} = \sum_\nu \mathcal{D}^\lambda_{\mu\nu}(\theta_1, \theta_2, \theta_3) a_{\lambda\nu}, \text{ rotationally invariant of } R \quad (3)$$

where $a_{\lambda\nu}$ are the deformation parameters in the body-fixed (intrinsic) coordinates, $\mathcal{D}^\lambda_{\mu\nu}$ is a rotational operator. In quadrapole deformation, if we chose the principal coordinates as the body fixed coordinates then one can easily prove that $a_{21} = a_{2,-1} = 0$, $a_{22} = a_{2,-2} \neq 0$, $a_{2,0} \neq 0$. That is we need only two coordinates $a_0$ and $a_2$ (for abbreviation we drop the subscript 2 which represents $\lambda$) to specify the shape of the nucleus and we need the three Euler angles $(\theta_1, \theta_2, \theta_3)$ to specify the orientation of the principal axes with respect to the space fixed coordinates.

Sometimes it is more convenient to use Bohr notation $(\beta, \gamma)$ instead of $a_0$ and $a_2$ with

$$a_0 = \beta \cos(\gamma) \quad (4a)$$

$$a_2 = \frac{\beta}{\sqrt{2}} \sin(\gamma) \quad (4b)$$

The hydrodynamical model assumes an irrotational flow for the nuclear matter $(\boldsymbol{\nabla} \times \mathbf{v}(\mathbf{r}) = 0)$ and the incompressibility $\boldsymbol{\nabla}.\mathbf{v}(\mathbf{r}) = 0$ and hence the velocity can be derived from a scalar potential $\chi(\mathbf{r})$ as

$$\mathbf{v}(\mathbf{r}) = \boldsymbol{\nabla}\chi(\mathbf{r})$$

According to the assumption above, $\chi(\mathbf{r})$ is the general solution of the Laplace equation $\nabla^2\chi(\mathbf{r}) = 0$. That is

$$\chi(\mathbf{r}) = \sum_\mu \xi^*_{2\mu} r^2 Y_{2\mu}(\theta, \varphi) \quad (5)$$

where $\xi^*_{2\mu}$ is parameters can be determined from the boundary conditions. If the velocity at the surface of the nucleus is assumed to be radial then

$$v_{r=R} = \left(\frac{\partial \chi}{\partial r}\right)_{r=R} = \frac{dR}{dt} \quad (6)$$

The left side of (6) gives

$$v_{r=R} = \sum_\mu 2R\xi^*_{2\mu} Y_{2\mu}(\theta, \varphi) \quad (7a)$$

And the right hand side is

$$v_{r=R} = R_0 \sum_\mu \dot{\alpha}^*_{2\mu} Y_{2\mu}(\theta, \varphi) \quad (7b)$$

For small oscillation $R$ in (7a) can be approximated to $R_0$ and then by equating both equations after replacing $R$ in (7a) by $R_0$ one can easily obtain

$$\xi^*_{2\mu} = \frac{1}{2}\dot{\alpha}^*_{2\mu} \quad (8)$$

The kinetic energy of the entire liquid drop is given by

$$T = \frac{1}{2}\rho_0 \int d\tau v^2 = \frac{1}{2}\rho_0 \frac{1}{4}\sum_{\mu\mu'} \dot{\alpha}^*_{2\mu}\dot{\alpha}^*_{2\mu'} \int d\omega \int_0^{R(\theta,\varphi)} r^2 dr \nabla \cdot \left(r^2 Y_{2\mu}(\theta,\varphi)\right) \cdot \nabla\left(r^2 Y_{2\mu'}(\theta,\varphi)\right) \quad (9)$$

The second line in (9) is obtained by using of the eq. (5) and (8). The definition of $\boldsymbol{\nabla}$ is

$$\boldsymbol{\nabla} = \left(\frac{\partial}{\partial r}, \frac{1}{r}\frac{\partial}{\partial \theta}, \frac{1}{r sin\theta}\frac{\partial}{\partial \varphi}\right), d\tau = r^2 sin\theta dr d\theta d\varphi = r^2 dr d\omega.$$

Putting this definition in (9) it follows

$$T =$$
$$\frac{1}{8}\rho_0 \sum_{\mu\mu'} \dot{\alpha}^*_{2\mu}\dot{\alpha}^*_{2\mu'} \int d\omega \left\{\left[4Y_{2\mu}Y_{2\mu'} + \frac{\partial Y_{2\mu}}{\partial\theta}\frac{\partial Y_{2\mu'}}{\partial\theta} + cosec^2\theta \frac{\partial Y_{2\mu}}{\partial\varphi}\frac{\partial Y_{2\mu'}}{\partial\varphi}\right] \int_0^{R(\theta\varphi)} r^4 dr\right\} \tag{10}$$

The integration over $r$ in (10) gives

$$\int_0^{R(\theta,\varphi)} r^4 dr = \frac{R^5}{5} = \frac{R_0^5}{5}(1+\varepsilon)^5$$

$$= \frac{R_0^5}{5}(1 + 5\varepsilon + 10\varepsilon^2 + \cdots). \tag{11}$$

where $\varepsilon = \sum_{\lambda\mu} \alpha^*_{\lambda\mu} Y_{\lambda\mu}(\theta,\varphi)$. In (11) we use Taylor expansion $(x) = \sum_r f^r(0)x^r/r!$ , where $f^r(0)$ is the $r$ – differential of $f$ with respect to $x$ at $x$ = 0. In the case of charged liquid drop Rayleigh assumes the surface oscillations are small and because of this he consider only the first term of equation (11). In our case the first three terms are treated.

Substitution by (11) in (10) and using the identities

$$\frac{\partial}{\partial\theta} = \frac{1}{2}\left(L_+e^{-i\varphi} - L_-e^{+i\varphi}\right)$$

$$i\, cot\theta \frac{\partial}{\partial\varphi} = \frac{1}{2}\left(L_+e^{-i\varphi} + L_-e^{+i\varphi}\right),$$

eq. (10) becomes

$$T = \frac{1}{8}\rho_0 \sum_{\mu\mu'} \dot{\alpha}^*_{2\mu}\dot{\alpha}^*_{2\mu'} \int d\omega \left\{\left[(4-\mu\mu')Y_{2\mu}Y_{2\mu'} - L_+Y_{2\mu}L_-Y_{2\mu'}\right]\frac{R_0}{5}(1+5\varepsilon+10\varepsilon\cdots)\right\} \tag{12}$$

## B. Working out the integration in (1.12) with the first term of R- expansion (Bohr calculations)

This part was worked out by Bohr[11] who consider only the first term in eq. (11). That is he regard the change in $R$ due to the deformation is small to an extent that he can ignore all other terms in that equation

The equation of kinetic energy with the first term only is written as

$$T = \frac{1}{8}\rho_0 \sum_{\mu\mu'} \dot{\alpha}^*_{2\mu}\dot{\alpha}^*_{2\mu'} \int d\omega \left\{\left[(4-\mu\mu')Y_{2\mu}Y_{2\mu'} - L_+Y_{2\mu}L_-Y_{2\mu'}\right]\frac{R_0^5}{5}\right\} \tag{13}$$

By the use of the identity[12, 13]

$$L_\pm Y_{\lambda\mu} = \sqrt{(\lambda\mp\mu)(\lambda\pm\mu+1)}\,Y_{\lambda\mu\pm1} \tag{14}$$

and the orthogonal property of the spherical harmonics $\int Y_{2\mu} Y_{2\mu'} d\omega = \int Y_{2\mu} Y^*_{2,-\mu'} (-1)^{\mu'} d\omega = (-1)^{\mu} \delta_{\mu,-\mu'}$, eq. (13) becomes

$$T = \frac{1}{8}\rho_0 \sum_{\mu\mu'} \dot{\alpha}^*_{2\mu} \dot{\alpha}^*_{2\mu'} (10 - \mu) \qquad (15)$$

Because $\mu$ runs from -2 to 2, the summation over $\mu$ is zero and (15) becomes

$$T = \frac{1}{2}\sum_{\mu} B_2 |\dot{\alpha}_{2\mu}|^2 \text{ , with } B_2 = \frac{\rho_0 R_0^5}{2} \qquad (16)$$

Eq. (16) represents the kinetic energy in space fixed coordinate. From eq. (3) we have

$$\dot{\alpha}_{\lambda\mu} = \sum_{v} \mathcal{D}^{\lambda}_{\mu v} \dot{a}_{\lambda v} + \dot{\mathcal{D}}^{\lambda}_{\mu v} a_{\lambda v} \qquad (17)$$

Inserting this result in (16) the kinetic energy is given by

$$T = \frac{1}{2} B \sum_{\mu} \sum_{vv'} \left( \mathcal{D}^{2*}_{\mu v} \mathcal{D}^{2}_{\mu v'} \dot{a}^*_{2v} \dot{a}_{2v'} + \dot{\mathcal{D}}^{2*}_{\mu v} \dot{\mathcal{D}}^{2}_{\mu v'} a^*_{2v} a_{2v} + \dot{\mathcal{D}}^{2*}_{\mu v} \mathcal{D}^{2}_{\mu v} \dot{a}^*_{2v} a_{2v} \right) \qquad (18)$$

The three parts in (18) are respectively, vibrational, rotational and cross terms. The cross terms was proved zero and $\dot{\mathcal{D}}$ in the second term of eq. (18) is the time derivative of the rotational transformation matrix. It is given by[14]

$$\dot{\mathcal{D}}^{2}_{\mu v}(\theta_k) = -i \sum_{v',k} \mathcal{D}^{\lambda}_{\mu v'}(\theta_k) \langle 2\ v' | L_k | 2\ v \rangle\, \omega_k \qquad (19)$$

Inserting (19) in (18) and cancelling the third term from (18) yields

$$T = \frac{1}{2} B \sum_{\mu} \sum_{vv'} \left( \mathcal{D}^{2*}_{\mu v} \mathcal{D}^{2}_{\mu v'} \dot{a}^*_{2v} \dot{a}_{2v'} + \sum_{\substack{v'' v''' \\ k,k'}} \mathcal{D}^{2*}_{\mu v''} \mathcal{D}^{\lambda}_{\mu v'''} \langle 2\ v'' | L_k | 2\ v \rangle^* \langle 2\ v''' | L_{k'} | 2\ v' \rangle a^*_v a_v \omega_k \omega_{k'} \right) , \qquad (20)$$

in the second part of eq. (20), $v, v', ..$ can only take the even values $0, 2, -2$ because of this $k$ and $k'$ cannot be different. By the use of the unitary property of the transformation matrix $\mathcal{D}^{2*}_{\mu v''}$ , (i. e., $\sum_{v'' v'''} \mathcal{D}^{2*}_{\mu v''} \mathcal{D}^{\lambda}_{\mu v'''} = \delta_{v'', v'''}$,) eq. (20) can be rewritten as

$$T = \frac{1}{2} B \left( \sum_{\mu v} |\dot{a}_v|^2 + \sum_{vv'} \sum_{k} \langle 2v' | L^2_k | 2v \rangle a^*_v a_{v'} \omega^2 \right) \qquad (21)$$

If we use the Bohr notation we should differentiate eq. (1.4) with respect to time taking in our account $\gamma$ approaches zero in the special case of axially symmetric nuclei

$$a_0 = \beta cos\gamma \Rightarrow \dot{a}_0 = \dot{\beta} cos\gamma - \beta\dot{\gamma} \sin\gamma = \dot{\beta} \qquad (22a)$$

$$a_2 = \frac{\beta}{\sqrt{2}} \sin\gamma \Rightarrow \dot{a}_2 = \frac{\dot{\beta}}{\sqrt{2}} \sin\gamma + \frac{\beta}{\sqrt{2}} \dot{\gamma} \cos\gamma = \frac{\beta}{\sqrt{2}} \dot{\gamma} \qquad (22b)$$

Putting these two equations in the first part (vibrational part) of eq. (21) gives

$$\frac{1}{2}B\sum_{\mu\nu}|\dot{a}_\nu|^2 = \frac{1}{2}B(|\dot{a}_0|^2 + |\dot{a}_2|^2 + |\dot{a}_{-2}|^2) = \frac{1}{2}B(\dot{\beta}^2 + \beta^2\dot{\gamma}^2) \qquad (23)$$

The second part of equation (21) is the rotational part. That is

$$T_{rot} = \frac{1}{2}B\sum_{\nu\nu'}\sum_k \langle 2\nu\,|L_k^2|2\,\nu'\rangle\, a_\nu^* a_{\nu'}\,\omega^2 = \frac{1}{2}\sum_k \mathcal{J}_k\omega^2 \qquad (24)$$

It follows

$$\mathcal{J}_k = B\sum_{\nu\,\nu'}\langle 2\,\nu'|L_k^2|2\,\nu\rangle a_\nu a_{\nu'}\;. \qquad (25)$$

Eq. (25) represents the moment of inertia $\mathcal{J}_k$ along the axis $k$ which arises from the first term in (11), we will call it the contribution of the first term in (11) or the formula of Bohr, let us denote it by $\mathcal{J}_{Bk}$. The main aim of this work is to discuss the effect of the second and third terms in (11) separately on the value of $\mathcal{J}_k$.

## C. working out the integration in (1.12) with the second term in R- expansion

Where Eq. (12) with the second term can be written as

$$T'_{Rot} = \frac{1}{8}\rho_0 R_0^5 \sum_{\mu\mu'}\dot{\alpha}_\mu^*\dot{\alpha}_{\mu'}^*\sum_{\mu''}\alpha_{\mu''}^*\int d\omega\{[(4-\mu\mu')Y_{2\mu}Y_{2\mu'} - L_+Y_{2\mu}L_-Y_{2\,\mu'}]Y_{2\,\mu''}\} \qquad (26a)$$

$$= \frac{1}{8}\rho_0 R_0^5\sum_\mu|\dot{\alpha}_{2\mu}|^2\sum_{\mu''}\alpha_{\mu''}^*\int d\omega\{Y_{2\mu''}[(4+\mu^2)Y_{2\mu}Y_{2\mu}^* + (2-\mu)(2+\mu+1)Y_{2\,\mu+1}Y_{2\,\mu+1}^*]\} \qquad (26b)$$

$$= \frac{1}{8}\rho_0 R_0^5\sum_\mu|\dot{\alpha}_{2\mu}|^2\alpha_{2\mu''}^*(10+\mu)\int d\omega Y_{2\,\mu''}Y_{2\mu}Y_{2\mu}^* \qquad (26c)$$

$$= \frac{1}{8}\rho_0 R_0^5\sum_\mu|\dot{\alpha}_{2\mu}|^2\alpha_{20}^*(10+\mu)\left(\frac{5}{4\pi}\right)^{1/2}\langle 220\mu|2\mu\rangle\langle 2200|20\rangle \qquad (26d)$$

$$= CB\sum_\mu|\dot{\alpha}_{2\mu}|^2\alpha_{20}^*\langle 220\mu|2\mu\rangle \qquad (26e)$$

where, $C = \frac{10}{4}\left(\frac{5}{4\pi}\right)^{1/2}\langle 2200|20\rangle$ and $B = \frac{1}{2}\rho_0 R_0^5$. In getting the first line (Eq. 26a) we have used the definition of $\varepsilon$. In second line we use the identity $L_\pm Y_{2\,\mu} = \sqrt{(2\mp\mu)(2\pm\mu+1)}Y_{2\,\mu\pm 1}$ and the orthogonal property the spherical harmonics as we do in part A. Since the subscript $\mu+1$ in Eq. (26b) is dummy variable, it can be replaced by $\mu$ without any change in the value of integration (this idea is checked also by Maple). The integration in (26c) on $Y'$s can be nonzero only when $\mu''=0$. In getting the fourth line from the third line we used the identity:

$$\int d\omega Y_{\lambda_1\mu_1}Y_{\lambda_2\mu_2}Y^*_{\lambda_3\mu_3}=\left(\frac{(2\lambda_1+1)(2\lambda_2+1)}{4\pi(2\lambda_3+1)}\right)^{1/2}\langle\lambda_1\lambda_2\mu_1\mu_2|\lambda_3\mu_3\rangle\langle\lambda_1\lambda_2 00|\lambda_3 0\rangle$$

We use the symbol $T'_{Rot}$to indicate that the concentration is only on rotational part. Eq. (26e) represents the first correction in kinetic energy in space fixed coordinate. To transform it to the body fixed coordinates one should use the transformation relation $\alpha_{\lambda\mu}=\sum_\nu \mathcal{D}^{\lambda}_{\mu\nu}(\theta_k)a_{\lambda\nu}$, that was used in part (1). In this part we will concentrate only on rotational part (i.e. the part contains $\dot{D}$)

$$\begin{aligned}T'_{rot}&=CB\sum_{\mu}\sum_{\nu\nu'\sigma'}\dot{D}^{2*}_{\mu\nu}\dot{D}^{2}_{\mu\nu'}D^{2}_{0\sigma}a_\nu a_{\nu'}a_\sigma\langle 220\mu|2\mu\rangle\\&=CB\sum_{\nu\nu'\sigma'}\sum_{k\nu''k'\nu'''}\underbrace{\sum_{\mu}D^{2*}_{\mu\nu''}D^{2}_{\mu\nu'''}D^{2}_{0\sigma}a_\nu a_{\nu'}a_\sigma\langle 220\mu|2\mu\rangle}_{=\langle 22\nu'''|2\nu''\rangle}\langle 2\nu''|L_k|2\nu'''\rangle\langle 2\nu''|L_{k'}|2\nu'''\rangle^{*}a_\nu a_{\nu'}a_\sigma\omega_k\omega_{k'}\end{aligned}\quad(27)$$

In the second line we have used the definition of $\dot{D}^2_{\mu\nu'}$ in (19). Using the conditions that $k=k',\nu'''=\nu''$ for the same reason mentioned in part B and hence $\sigma=0$ the equation of $T'$ becomes:

$$T'_{Rot}=CB\sum_{\nu\nu'}\sum_{k\nu''}\langle 22\nu''0|2\nu''\rangle\langle 2\,\nu|L_k|2\nu''\rangle\langle 2\,\nu''|L_k|2\,\nu'\rangle a_\nu a_{\nu'}a_0\omega_k^2=\frac{1}{2}\sum_k \mathcal{J}'_k\omega_k^2\quad(28)$$

where the quantities $\mathcal{J}'_k$ are the first correction to moment of inertia are given by:

$$\mathcal{J}'_k=2CB\sum_{\nu\nu'}\sum_{\nu''}\langle 22\nu''0|2\nu''\rangle\langle 2\nu'|L_k|2\nu''\rangle\langle 2\,\nu''|L_k|2\nu\rangle a_\nu a_{\nu'}a_0$$

$$=2CB\sum_{\nu\nu'}\sum_{\nu''}\langle 22\nu''0|2\nu''\rangle\langle 2\nu'|L_k^2|2\nu\rangle a_\nu a_{\nu'}a_0\quad(29)$$

### D. The integration over the third term in (12)

where eq. (12) with the third term is written as

$$T''=\frac{1}{8}\rho_0\sum_{\mu\mu'}\dot{\alpha}^*_{2\mu}\dot{\alpha}^*_{2\mu'}\int d\omega\left\{[(4-\mu\mu')Y_{2\mu}Y_{2\mu'}-L_+Y_{2\mu}L_-Y_{2\mu'}]\frac{R_0^5}{5}(10\varepsilon^2)\right\}$$

$$=\frac{1}{8}\rho_0\sum_{\mu\mu'}\dot{\alpha}^*_{2\mu}\dot{\alpha}^*_{2\mu'}\sum_{\sigma\sigma'}\alpha^*_\sigma\alpha^*_{\sigma'}\int d\omega\{[(4-\mu\mu')Y_{2\mu}Y_{2\mu'}-L_+Y_{2\mu}L_-Y_{2\mu'}]Y_{2\sigma}Y_{2\sigma'}\}$$

$$= \frac{1}{8}\rho_0 \sum_{\mu\mu'} \dot{\alpha}^*_{2\mu}\dot{\alpha}^*_{2\mu'} \sum_{\sigma\sigma'} \alpha^*_{\sigma}\alpha^*_{\sigma'} \int d\omega \left\{\left[(4-\mu\mu')Y_{2\mu}Y_{2\mu'} - \sqrt{(2-\mu)(2+\mu+1)}\sqrt{(2+\mu')(2-\mu'+1)}\; Y_{2\mu+1}Y_{2\mu'-1}\right]Y_{2\sigma}Y_{2\sigma'}\right\} \quad (30)$$

In the second line of eq. (D.1) we use the value $\varepsilon = \sum_{\lambda\mu} \alpha^*_{\lambda\mu} Y_{\lambda\mu}(\theta,\varphi)$ , in third line the identity $L_{\pm}Y_{\lambda\mu} = \sqrt{(\lambda\mp\mu)(\lambda\pm\mu+1)}Y_{\lambda\mu\pm1}$ is used. As we did in part B, because of the orthogonality of $\dot{\alpha}$ and $\alpha$ we can write the summation over $\mu$ and $\mu'$ as $\sum_{\mu\mu'} \dot{\alpha}^*_{2\mu}\dot{\alpha}^*_{2\mu'} = \sum_{\mu\mu'} \left|\dot{\alpha}_{2\mu}\right|^2(-1)^{\mu}\,\delta_{\mu,-\mu'}$ and the same for the summation over $\sigma$ and $\sigma'$ then eq. (3.1) becomes

$$T'' = \frac{1}{8}\rho_0 \sum_{\mu\mu'} \left|\dot{\alpha}_{2\mu}\right|^2 \sum_{\sigma\sigma'} |\alpha_{2\sigma}|^2 \int d\omega \left\{\left[(4-\mu\mu')(-1)^{\mu}\delta_{\mu,-\mu'}Y_{2\mu}Y_{2\mu'} - (-1)^{\mu}\delta_{\mu,-\mu'}\sqrt{(2-\mu)(2+\mu+1)}\sqrt{(2+\mu')(2-\mu'+1)}\; Y_{2\mu+1}Y_{2\mu'-1}\right](-1)^{\sigma}\delta_{\sigma,-\sigma'}Y_{2\sigma}Y_{2}\right.$$

$$= \frac{1}{8}\rho_0 \sum_{\mu} \left|\dot{\alpha}_{2\mu}\right|^2 \sum_{\sigma} |\alpha_{2\sigma}|^2 \int d\omega\left\{\left[(4+\mu^2)(-1)^{\mu}Y_{2\mu}Y_{2,-\mu} - (-1)^{\mu}(2-\mu)(2+\mu+1)\; Y_{2\mu+1}Y_{2,-(\mu+1)}\right](-1)^{\sigma}Y_{2\sigma}Y_{2,-\sigma}\right\} \quad (31)$$

Again because the subscript $(\mu+1)$ is dummy variable it can be replaced by $\mu$

$$T'' = \frac{1}{8}\rho_0 \sum_{\mu} \left|\dot{\alpha}_{2\mu}\right|^2 \sum_{\sigma} |\alpha_{2\sigma}|^2 \int d\omega\left\{(10+\mu)(-1)^{\mu}Y_{2\mu}Y_{2,-\mu}(-1)^{\sigma}Y_{2\sigma}Y_{2,-\sigma}\right\} \quad (32)$$

using the identity

$$Y_{j_1m_1}Y_{j_2m_2} = \left(\frac{2j_1+1}{4\pi}\right)^{1/2}\left(\frac{2j_2+1}{4\pi}\right)^{1/2}\sum_j\left(\frac{4\pi}{2j+1}\right)^{1/2} Y_{jm}\langle j_1j_2m_1m_2|jm\rangle\langle j_1j_2 00|j0\rangle \quad (33)$$

eq (30) becomes

$$T'' = \frac{1}{2}B\sum_{\mu}\left|\dot{\alpha}_{2\mu}\right|^2\sum_{\sigma}|\alpha_{2\sigma}|^2\frac{250}{4\pi}\int(-1)^{\mu}(-1)^{\sigma}\sum_{jj'}Y_{j0}Y_{j'0}\langle j_1j_2\mu-\mu|j0\rangle\langle j_1j_2 00|j0\rangle\langle j_1j_2\sigma-\sigma|j0\rangle\langle j_1j_2 00|j0\rangle$$

$$= \frac{1}{2}B\sum_{\mu}\left|\dot{\alpha}_{2\mu}\right|^2\sum_{\sigma}|\alpha_{2\sigma}|^2\frac{250}{4\pi}\int \underbrace{\sum_j\langle j_1j_2\mu-\mu|j0\rangle\langle j_1j_2 00|j0\rangle Y_{j0}(-1)^{\mu}}_{Y_{00}}\;\underbrace{\sum_{j'}\langle j_1j_2\sigma-\sigma|j'0\rangle\langle j_1j_2 00|j'0\rangle(-1)^{\sigma}Y_{j'0}}_{=Y_{00}}$$

$$T'' = \frac{1}{2}B\sum_{\mu}\left|\dot{\alpha}_{2\mu}\right|^2\sum_{\sigma}|\alpha_{2\sigma}|^2\frac{250}{4\pi} \quad (34)$$

In (24), the quantity $\frac{1}{2}B\sum_{\mu}|\dot{\alpha}_{2\mu}|^2$ in body fixed frame is already treated in part A. treated before to be equal to. The summation over $\sigma$ equals $\beta^2$ and

$$T'' = \frac{250}{4\pi}\beta^2\frac{1}{2}B\sum_{\nu\nu'k}\langle 2\ \nu'|L_k|2\ \nu\rangle a_\nu a_{\nu'}\omega_k^2 = \frac{1}{2}\sum_k \mathcal{J}_k''\omega_k^2 \quad (35)$$

That is

$$\mathcal{J}_k'' = \frac{250}{4\pi}\beta^2 \underbrace{B\sum_{\nu\nu'k}\langle 2\ \nu'|L_k|2\ \nu\rangle a_\nu a_{\nu'}}_{=(\mathcal{J}_k)_{Bohr}} \quad (36)$$

or

$$\mathcal{J}_k'' = \frac{250}{4\pi}\beta^2\mathcal{J}_k \quad (37)$$

Where $\mathcal{J}_k''$ is the second correction to the moment of inertia, $\mathcal{J}_k$ is the uncorrected one

## E. Calculation and Discussions

a. The contribution of the first term only

In order to discuss the effect of the first tem we should first to extract the values of $\mathcal{J}_1, \mathcal{J}_2, \mathcal{J}_3$ from (25). In order to do so we use the identities[13, 15]

$$L_1 = \frac{1}{2}(L_+ + L_-) \Rightarrow L_1^2 = \frac{1}{4}(L_+^2 + L_-^2 + 2L_+L_-) \quad (38a)$$

$$L_2 = \frac{1}{2i}(L_+ - L_-) \Rightarrow L_2^2 = \frac{-1}{4}(L_+^2 + L_-^2 - 2L_+L_-) \quad (38b)$$

$$L_+L_- = L^2 - L_z^2 \quad (38c)$$

$$\langle 2\nu \pm 1\ |(L_\pm)|2\ \nu\rangle = \sqrt{(2 \mp \nu)(2 \pm \nu + 1)} \quad (38d)$$

By the use of (38)

$$\mathcal{J}_1 = B\sum_{\nu\nu'}\langle 2\ \nu'|L_1^2|2\ \nu\rangle a_\nu a_{\nu'}$$

$$= \frac{1}{4}B\sum_{\nu\nu'}\langle 2\ \nu'|L_+^2 + L_-^2 + 2L_+L_-|2\ \nu\rangle a_\nu a_{\nu'} \quad (39)$$

The working on the first term in (39)

$$\sum_{\nu\nu'}\langle 2\ \nu'|L_+^2|2\ \nu\rangle a_\nu a_{\nu'} = B\sum_{\nu}\langle 2\ \ \nu+2|L_+|2\ \ \nu+1\rangle\langle 2\ \nu+1|L_+|2\ \nu\rangle a_\nu a_{\nu+2}$$

$$= \{\langle 2\ 0|L_+|2\ \ -1\rangle\langle 2-1|L_+|2-2\rangle a_{-2}a_0 + \langle 2\ 2|L_+|2\ 1\rangle\langle 21|L_+|20\rangle a_0 a_2\}$$

$$= 4\sqrt{6}a_0a_2 \quad (40a)$$

Be note that $a_2$ and $a_{-2}$ are equal. We will do the same for the second term in eq. (1.23) and we get

$$\sum_{\nu\nu'}\langle 2\ \nu'|L_-^2|2\ \nu\rangle a_\nu a_{\nu'} = 4\sqrt{6}a_0a_2 \tag{40b}$$

The third term is very simple

$$\sum_{\nu\nu'}\langle 2\ \nu'|2L_+L_-|2\ \nu\rangle a_\nu a_{\nu'} = 2\sum_{\nu\nu'}\langle 2\ \nu'|L^2 - L_z^2|2\ \nu\rangle a_\nu a_{\nu'}$$

$$= 2\sum_{\nu}\langle 2\ \nu|(2(2+1)-\nu^2)|2\ \nu\rangle a_\nu a_\nu$$

$$= 2B\{(2(2+1)-(-2)^2)a_{-2}a_{-2} + (2(2+1)-(0)^2)a_0a_0 + (2(2+1)-(2)^2)a_2a_2\}$$

$$= 2\{4a_2a_2 + 6a_0a_0\} \tag{40c}$$

Inserting (40) in (39) one can obtain

$$\mathcal{J}_1 = B(2\sqrt{6}a_0a_2 + 2a_2^2 + 3a_0^2) \tag{41}$$

By the same method we can estimate that

$$\mathcal{J}_2 = B(-2\sqrt{6}a_0a_2 + 2a_2^2 + 3a_0^2) \tag{42}$$

Now we are going to find out $\mathcal{J}_3$

$$\mathcal{J}_3 = B\sum_\nu\langle 2\ \nu|L_3^2|2\ \nu\rangle a_\nu a_\nu$$

$$= B\{\langle 2-2|L_3^2|2-2\rangle a_{-2}a_{-2} + \langle 2\ 0|L_3^2|2\ \nu 0\rangle a_0a_0 + \langle 2\ 2|L_3^2|2\ 2\rangle a_2a_2\}$$

$$= B\{(-2)^2a_{-2}a_{-2} + (0)^2a_0a_0 + (2)^2a_2a_2\}$$

$$= 8Ba_2^2 \tag{43}$$

Eq.'s (41, 42, and 43) can be written in terms of Bohr notation of $a_0 = \beta\cos(\gamma)$, $a_2 = \frac{\beta}{\sqrt{2}}\sin(\gamma)$ in a compact form as:

$$\mathcal{J}_k = 4B\beta^2 sin^2\left(\gamma - k\frac{2\pi}{3}\right) \tag{44}$$

For axially symmetric nuclei where $\gamma = 0$

$$\mathcal{J}_1 = \mathcal{J}_2 = 3B\beta^2 \tag{45a}$$

$$\mathcal{J}_3 = 0 \tag{45b}$$

b. The contribution of the second term (the first part of this work)

In order to evaluate $\mathcal{J}_1', \mathcal{J}_2', \mathcal{J}_3'$ from eq. (37) we follow the same procedure that is used in the first part. Be careful that $\nu$ and $\nu'$ take only the values $-2, 0, 2$ and the values of $\nu''$ is controlled by this condition, also $a_2 = a_{-2}$

$$\mathcal{J}_1' = 2CB\sum_{vv'}\sum_{v''}\langle 22v''0|2v''\rangle\langle 2v'\,|L_1^2|2v\,\rangle a_v a_{v'} a_0$$

$$= \frac{1}{4} * 2CB\sum_{vv'}\sum_{v''}\langle 22v''0|2v''\rangle\,\langle 2v'\,|(L_+^2 + L_-^2 + 2L_+L_-)|2v\,\rangle a_v a_{v'} a_0 \qquad (46)$$

The working on $L_+^2$, in this case $v'' = v + 1$ and $v' = v + 2$

$$2CB\sum_{vv'}\sum_{v''}\langle 22v''0|2v''\rangle\langle 2v'\,|L_+^2|2v\,\rangle a_v a_{v'} a_0 = 2CB\sum_v\langle 220v+1|2v+1\rangle\langle 2v+2\,|L_+^2|2v\,\rangle a_v a_{v+2} a_0 \qquad (47)$$

$$= \langle 22-10|2-1\rangle\langle 20\,|L_+|2-1\,\rangle\langle 2-1|L_+|2-2\rangle a_2 a_0^2 + \langle 2210|21\rangle\langle 22\,|L_1|21\,\rangle\langle 21|L_1|20\rangle a_2 a_0^2$$

$$= -4BC\sqrt{\frac{2}{7}}\,\sqrt{6}\,a_2 a_0^2 \qquad (48a)$$

The same thing is doing for $L_-^2$. That is

$$2CB\sum_{vv'}\sum_{v''}\langle 22v''0|2v''\rangle\langle 2v'\,|L_-^2|2v\,\rangle a_v a_{v'} a_0 = -4BC\sqrt{\frac{2}{7}}\,\sqrt{6}\;a_2 a_0^2 \qquad (48b)$$

In the third term $v'' = v' = v$

$$2CB\sum_{vv'}\sum_{v''}\langle 22v''0|2v''\rangle\langle 2v'\,|2L_+L_-|2v\,\rangle a_v a_{v'} a_0 = 2 * 2CB\sum_{vv'}\sum_{v''}\langle 22v''0|2v''\rangle\,\langle 2v'|L_+L_-|2v\,\rangle a_v a_{v'} a_0 = 2 * 2CB\sum_{vv'}\sum_{v''}\langle 22v''0|2v''\rangle\,\langle 2v'|L^2 - L_z^2|2v\,\rangle a_v a_{v'} a_0 = 2 * 2CB\sum_v\langle 22v0|2v\rangle\langle 2v|L^2 - L_z^2|2v\,\rangle a_v a_v a_0 = 2 * 2CB\sum_v\langle 22v0|2v\rangle\,(2(2+1) - v^2)a_v a_v a_0 = 2 * 2CB\{\langle 2220|22\rangle(2(2+1) - 2^2)a_2^2 a_0 + \langle 2200|20\rangle(2(2+1) - 0^2)a_0^3 + \langle 22-20|2-2\rangle(2(2+1) - (-2)^2)a_2^2 a_0\} = 2 * 2CB\left\{\sqrt{\frac{2}{7}}\,(2)a_2^2 a_0 - \sqrt{\frac{2}{7}}\,(6)a_0^3 + \sqrt{\frac{2}{7}}\,(2)a_2^2 a_0\right\} = 8CB\left\{\sqrt{\frac{2}{7}}\,(2)a_2^2 a_0 - \sqrt{\frac{2}{7}}\,(3)a_0^3\right\} \qquad (48c)$$

Putting eq. (C.7) in (C.6) we obtain

$$\mathcal{J}_1' = 2CB\sqrt{\frac{2}{7}}\{\,(2)a_2^2 a_0 - \,(3)a_0^3 - \,\sqrt{6}\,a_2 a_0^2\} \qquad (49)$$

By the same method $\mathcal{J}_2'$ can be calculated. The result is

$$\mathcal{J}_2' = 2CB\sqrt{\frac{2}{7}}\{\,(2)a_2^2 a_0 - \,(3)a_0^3 + \,\sqrt{6}\,a_2 a_0^2\} \qquad (50)$$

In calculating $\mathcal{J}_3'$ be noted that $v = v' = v''$

$$\mathcal{J}'_3 = 2CB\sum_{\nu} \langle 22\nu 0|2\nu\rangle\langle 2\nu\,|L_1^2|2\nu\,\rangle a_\nu^2 a_0 = 2CB\sum_{\nu} \langle 22\nu 0|2\nu\rangle \nu^2 a_\nu^2 a_0$$

$$= 2CB\{\langle 2220|22\rangle(2)^2 a_2^2 a_0 + \langle 2200|20\rangle(0)^2 a_0^2 a_0 + \langle 22-20|2-2\rangle \nu^2 a_{-2}^2 a_0\}$$

$$= 2CB\left\{\sqrt{\frac{2}{7}}(2)^2 a_2^2 a_0 - \sqrt{\frac{2}{7}}(0)^2 a_0^2 a_0 + \sqrt{\frac{2}{7}}(2)^2 a_{-2}^2 a_0\right\}$$

$$= 16\sqrt{\frac{2}{7}}CBa_2^2 a_0 \qquad (51)$$

The contribution of the third term (second correction)

The contribution of the third term or the (second correction) is simply given by the equation $\mathcal{J}''_k = \frac{250}{4\pi}\beta^2 \mathcal{J}_k$, where $\mathcal{J}_k$ are the values of the components of $\mathcal{J}$resulting from the contribution of the first term (Bohr contribution). In other words, the second correction is simply to multiply each component in $\mathcal{J}$ by a factor $\frac{250}{4\pi}\beta^2$.

## F. Summary

In summary, the contribution of each parts respectively are

a. In the case of triaxial nuclei

$$\mathcal{J}_{Bk} = 4B\beta^2 sin^2\left(\gamma - k\frac{2\pi}{3}\right),$$

$$\mathcal{J}'_1 = 2CB\sqrt{\frac{2}{7}}\{\,(2)a_2^2 a_0 - \,(3)a_0^3 - \sqrt{6}\,a_2 a_0^2\},$$

$$\mathcal{J}'_2 = 2CB\sqrt{\frac{2}{7}}\{\,(2)a_2^2 a_0 - \,(3)a_0^3 + \sqrt{6}\,a_2 a_0^2\},$$

$$\mathcal{J}'_3 = 16\sqrt{\frac{2}{7}}CBa_2^2 a_0,$$

$$\mathcal{J}''_k = \frac{250}{4\pi}\beta^2 \mathcal{J}_k = \frac{250}{\pi}Bsin^2\left(\gamma - k\frac{2\pi}{3}\right)\beta^4.$$

b. In the case of axial symmetry

$$\mathcal{J}_B = \mathcal{J}_{B1} = \mathcal{J}_{B2} = 3B\beta^2 \;,\quad \mathcal{J}_3 = 0,$$

$$\mathcal{J}' = \mathcal{J}'_1 = \mathcal{J}'_2 = -6CB\sqrt{\frac{2}{7}}a_0^3 = -6CB\sqrt{\frac{2}{7}}\beta^3 \;,\; \mathcal{J}'_3 = 0\,,$$

$$\mathcal{J}'' = \mathcal{J}''_1 = \mathcal{J}''_2 = \frac{750}{4\pi}B\beta^4 \;,\quad \mathcal{J}''_3 = 0.$$

And the total moment of inertia that results from the three contributions is

$$\mathcal{J}_{tot} = \mathcal{J}_B + \mathcal{J}' + \mathcal{J}'' = 3B\beta^2 - 6CB\sqrt{\frac{2}{7}}\beta^3 + \frac{750}{4\pi}B\beta^4 \qquad \text{(aa)}$$

In table (1) the deformation parameters[14] $\beta$ are extracted from the electric quadrupole transition $(E2)$, the experimental values of the moment of inertia (column 2 in the table) are deuced from the spacing in the ground state rotational bands[14]. Nuclei in this table are regarded as an axially symmetric so we use the equations in the first group b to evaluate Bohr values, the second and third line in the same group to find out the correc.1 and correc.2 in the table. The total theoretical value is the summation of all three parts. The last column is the total theoretical value multiplied by the factor $0.5/\beta$. All values of the moment of inertia are presented in terms of $\mathcal{J}_{rig} = B_{rig}(1 + 0.31\beta)$ where $B_{rig} = 0.0138A^{5/3}\hbar^2/MeV$, $\mathcal{J}_{rig}$ is the moment of inertia for a rigid body having the same volume, shape and density of the nucleus.

Two features are clearly shown in table (1); the values of the total moment of inertia comes from the contribution of the three terms in eq. (11) reach to nearly 0.7 of the experimental results which are much better than that of Bohr which not exceed 0.2 of the experimental values, this feature is shown more obviously in Fig (1) also which presents the comparison between the results of Bohr and that of this work relative to experimental ones. (2) the table shows also the contribution of each term separately, this enable us to find the contribution of each term separately and to define which one contributes more. The unusual conclusion is that the contribution comes from the third term is larger than that comes from the second term or even that of the first terms.

There is another feature that is revealed by fig (1). This feature is; when the experimental results are plotted against the deformation parameter $\beta$ the resulting graph looks like a band. The width of the band is very large. However, the curve that represent eq. (aa) after multiplied by the factor (0.5/ $\beta$) forms a lower limit of this band.

Table (1), the axially symmetric of the moment of inertia before and after corrections and after be multiplied by the factor $0.5/\beta$. All values of the moment of inertia are in terms of rigid body moment of inertia $\mathcal{J}_{rig}$ where $\mathcal{J}_{rig} = B_{rig}(1 + 0.31\beta$ and $B_{rig} = 0.0138A^{3/2}\hbar^2/MeV$ [16]

| nucleus | B | $\mathcal{J}/\mathcal{J}_{rig}$ | | | | | |
|---|---|---|---|---|---|---|---|
| | | Exp. | Bohr | Correc. 1 | Correc. 2 | Tot. Theo. | |
| $Sm^{150}$ | 0.184 | 0.147 | 0.029 | 0.005 | 0.019 | 0.053 | 0.143 |
| $Sm^{152}$ | 0.290 | 0.380 | 0.069 | 0.018 | 0.116 | 0.203 | 0.350 |
| $Sm^{154}$ | 0.336 | 0.551 | 0.092 | 0.028 | 0.206 | 0.325 | 0.483 |
| $Gd^{154}$ | 0.280 | 0.373 | 0.065 | 0.016 | 0.101 | 0.182 | 0.324 |
| $Gd^{156}$ | 0.320 | 0.498 | 0.083 | 0.024 | 0.170 | 0.277 | 0.433 |
| $Gd^{158}$ | 0.346 | 0.547 | 0.097 | 0.030 | 0.231 | 0.358 | 0.517 |
| $Gd^{160}$ | 0.354 | 0.561 | 0.101 | 0.032 | 0.252 | 0.385 | 0.544 |
| $Dy^{160}$ | 0.301 | 0.490 | 0.074 | 0.020 | 0.134 | 0.228 | 0.379 |
| $Dy^{162}$ | 0.320 | 0.512 | 0.083 | 0.024 | 0.170 | 0.277 | 0.433 |
| $Dy^{164}$ | 0.334 | 0.558 | 0.090 | 0.027 | 0.201 | 0.319 | 0.477 |
| $Er^{164}$ | 0.306 | 0.456 | 0.077 | 0.021 | 0.143 | 0.240 | 0.393 |
| $Er^{166}$ | 0.323 | 0.496 | 0.085 | 0.025 | 0.176 | 0.286 | 0.442 |
| $Er^{168}$ | 0.320 | 0.496 | 0.083 | 0.024 | 0.170 | 0.277 | 0.433 |
| $Er^{170}$ | 0.310 | 0.484 | 0.078 | 0.022 | 0.150 | 0.250 | 0.404 |
| $Yb^{170}$ | 0.304 | 0.455 | 0.076 | 0.021 | 0.139 | 0.235 | 0.387 |
| $Yb^{172}$ | 0.311 | 0.477 | 0.079 | 0.022 | 0.152 | 0.253 | 0.407 |
| $Yb^{174}$ | 0.308 | 0.475 | 0.078 | 0.022 | 0.146 | 0.245 | 0.398 |
| $Yb^{176}$ | 0.301 | 0.445 | 0.074 | 0.020 | 0.134 | 0.228 | 0.379 |
| $Hf^{176}$ | 0.30 | 0.41 | 0.074 | 0.020 | 0.132 | 0.226 | 0.376 |
| $Hf^{178}$ | 0.31 | 0.38 | 0.078 | 0.022 | 0.150 | 0.250 | 0.404 |
| $Hf^{180}$ | 0.27 | 0.38 | 0.060 | 0.015 | 0.087 | 0.162 | 0.300 |
| $W^{182}$ | 0.28 | 0.34 | 0.065 | 0.016 | 0.101 | 0.182 | 0.324 |
| $W^{184}$ | 0.25 | 0.31 | 0.052 | 0.012 | 0.065 | 0.128 | 0.256 |
| $W^{186}$ | 0.259 | 0.272 | 0.056 | 0.013 | 0.074 | 0.143 | 0.276 |
| $Os^{186}$ | 0.201 | 0.247 | 0.034 | 0.006 | 0.027 | 0.068 | 0.168 |
| $Os^{188}$ | 0.191 | 0.214 | 0.031 | 0.005 | 0.022 | 0.059 | 0.153 |
| $Os^{190}$ | 0.18 | 0.18 | 0.027 | 0.004 | 0.018 | 0.050 | 0.138 |
| $Os^{192}$ | 0.16 | 0.16 | 0.022 | 0.003 | 0.011 | 0.036 | 0.113 |
| $Pt^{194}$ | 0.152 | 0.0972 | 0.020 | 0.003 | 0.009 | 0.032 | 0.104 |
| $Pt^{196}$ | 0.122 | 0.0890 | 0.013 | 0.001 | 0.004 | 0.018 | 0.074 |
| $Pt^{198}$ | 0.13 | 0.076 | 0.015 | 0.002 | 0.005 | 0.021 | 0.081 |
| $Ra^{222}$ | 0.184 | 0.223 | 0.029 | 0.005 | 0.019 | 0.053 | 0.143 |
| $Ra^{224}$ | 0.171 | 0.291 | 0.025 | 0.004 | 0.014 | 0.043 | 0.126 |
| $Ra^{226}$ | 0.197 | 0.351 | 0.033 | 0.006 | 0.025 | 0.064 | 0.162 |
| $Ra^{228}$ | 0.212 | 0.400 | 0.038 | 0.007 | 0.034 | 0.079 | 0.186 |
| $Th^{226}$ | 0.220 | 0.330 | 0.041 | 0.008 | 0.039 | 0.088 | 0.199 |
| $Th^{228}$ | 0.225 | 0.403 | 0.042 | 0.009 | 0.043 | 0.094 | 0.208 |
| $Th^{230}$ | 0.233 | 0.433 | 0.045 | 0.010 | 0.049 | 0.104 | 0.223 |
| $Th^{232}$ | 0.243 | 0.450 | 0.049 | 0.011 | 0.058 | 0.118 | 0.242 |

| | | | | | | | | |
|---|---|---|---|---|---|---|---|---|
| $Th^{234}$ | 0.233 | 0.467 | 0.045 | 0.010 | 0.049 | 0.104 | 0.223 |
| $U^{230}$ | 0.245 | 0.443 | 0.050 | 0.011 | 0.060 | 0.121 | 0.246 |
| $U^{232}$ | 0.257 | 0.470 | 0.055 | 0.013 | 0.072 | 0.139 | 0.271 |
| $U^{234}$ | 0.251 | 0.516 | 0.052 | 0.012 | 0.066 | 0.130 | 0.258 |
| $U^{236}$ | 0.263 | 0.485 | 0.057 | 0.014 | 0.079 | 0.150 | 0.284 |
| $U^{238}$ | 0.268 | 0.480 | 0.059 | 0.014 | 0.085 | 0.159 | 0.296 |
| $Pu^{238}$ | 0.271 | 0.493 | 0.061 | 0.015 | 0.089 | 0.164 | 0.303 |
| $Pu^{240}$ | 0.278 | 0.488 | 0.068 | 0.018 | 0.111 | 0.196 | 0.342 |

Fig (1), represents the relation between the relative moment of inertia with respect to rigid one versus the deformation parameter β. the solid line represent Bohr, the dashed one is this work, the dashed-point curve is this work after multiplied by the factor 0.5/β

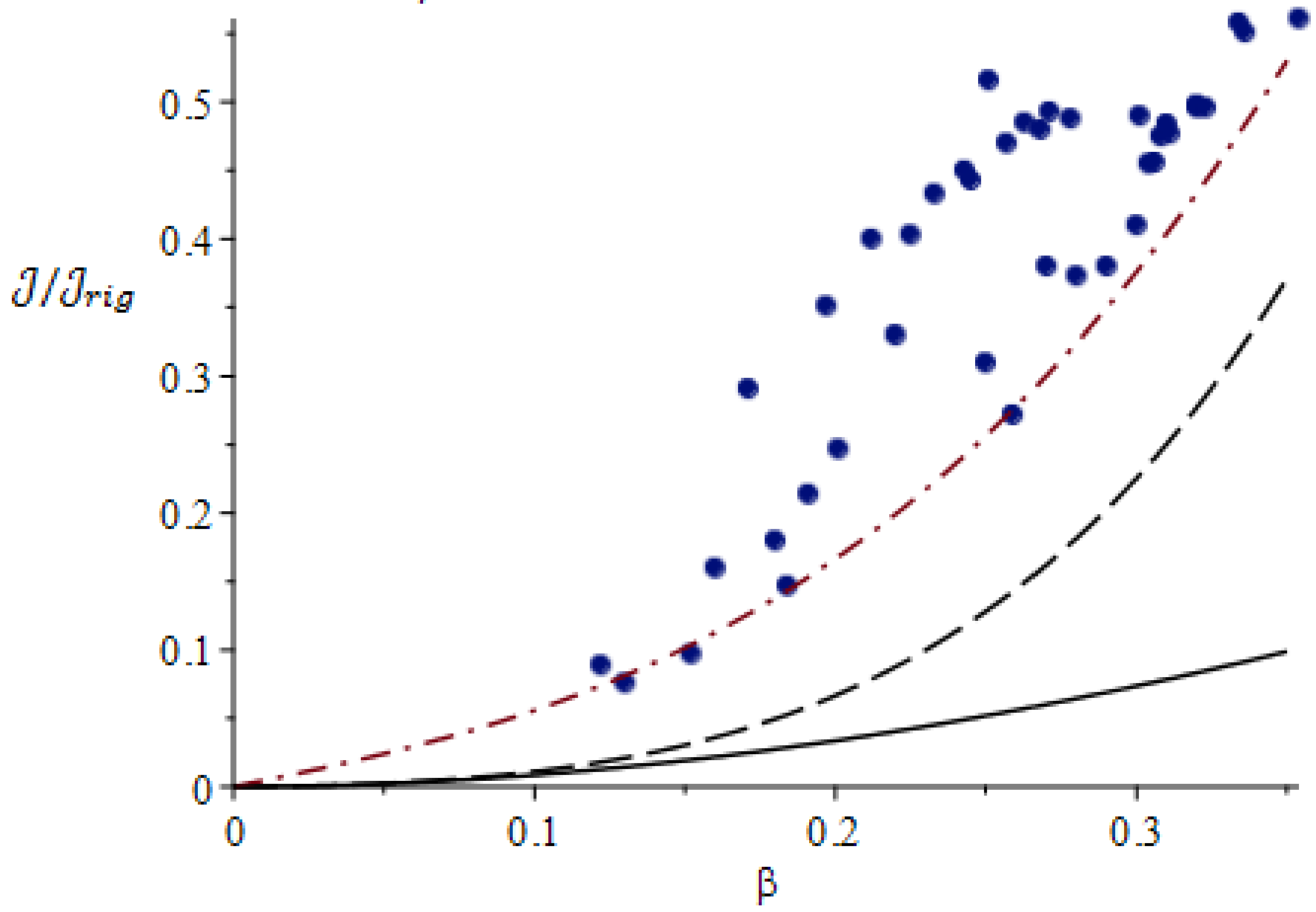

In this graph; the blue curve represents the Bohr results, the green one is our results the red one is our results after they are multiplied by the factor $0.5/\beta$.

Table 3: Comparative values of tri-axial momentum of inertia with the results from Bohr and experimental values respectively

| A | $\gamma$ | $\beta$ | $\mathfrak{J}_{B1}$ | $\mathfrak{J}_1$ Before | $\mathfrak{J}_1$ After | $\mathfrak{J}_1$ Exp. | $\mathfrak{J}_{B2}$ | $\mathfrak{J}_2$ Before | $\mathfrak{J}_2$ After | $\mathfrak{J}_2$ Exp. | $\mathfrak{J}_{B3}$ | $\mathfrak{J}_3$ Before | $\mathfrak{J}_3$ After | $\mathfrak{J}_3$ Exp. |
|---|---|---|---|---|---|---|---|---|---|---|---|---|---|---|
| 110 | 29 | **0.283** | 3.33 | 9.164 | 16.191 | 22 | 0.88 | 2.540 | 4.488 | 8.1 | 0.78 | 1.858 | 3.282 | 3.88 |
| 150 | 10.4 | **0.283** | 4.96 | 13.982 | 24.703 | 27.5 | 3.24 | 9.307 | 16.444 | 19.8 | 0.18 | 0.427 | 0.754 | 1.96 |
| 156 | 7.9 | **0.33** | 6.97 | 23.952 | 36.290 | 55 | 5.05 | 17.609 | 26.681 | 21.2 | 0.15 | 0.440 | 0.667 | 1.78 |
| 166 | 9.2 | **0.346** | 8.65 | 31.660 | 45.752 | 42.2 | 5.94 | 22.096 | 31.930 | 33.3 | 0.25 | 0.778 | 1.124 | 2.64 |
| 168 | 8.4 | **0.345** | 8.67 | 31.666 | 45.893 | 42.6 | 6.16 | 22.818 | 33.070 | 33.6 | 0.21 | 0.656 | 0.950 | 2.52 |
| 172 | 4.9 | **0.331** | 7.88 | 27.281 | 41.209 | 38.3 | 6.46 | 22.567 | 34.088 | 37.9 | 0.07 | 0.202 | 0.305 | 1.39 |
| 182 | 10 | **0.241** | 4.94 | 11.602 | 24.071 | 35.9 | 3.28 | 7.852 | 16.290 | 25.7 | 0.17 | 0.328 | 0.680 | 1.64 |
| 184 | 11.3 | **0.234** | 4.82 | 10.955 | 23.409 | 30.6 | 3.03 | 7.034 | 15.030 | 24.1 | 0.21 | 0.388 | 0.830 | 2.31 |
| 186 | 20.4 | **0.207** | 4.16 | 8.286 | 20.015 | 32.4 | 1.74 | 3.587 | 8.664 | 16.3 | 0.52 | 0.873 | 2.108 | 2.78 |
| 188 | 19.9 | **0.193** | 3.67 | 6.872 | 17.802 | 26.5 | 1.57 | 3.044 | 7.885 | 15.1 | 0.44 | 0.692 | 1.794 | 3.45 |
| 190 | 22.1 | **0.184** | 3.44 | 6.163 | 16.747 | 24.1 | 1.32 | 2.459 | 6.683 | 11.7 | 0.50 | 0.754 | 2.049 | 4.08 |

The moment of inertia $\mathfrak{J}_1, \mathfrak{J}_2, \mathfrak{J}_3$ of atomic nuclei with $E(4_1^+)/E(2_1^+) > 2.7$ that are extracted experimentally by Allmond and Wood[16] from $2_{g\gamma}^+$ energies and electric quadrupole matrix elements determined from multistep Coulomb excitation data are shown in table (2). For comparison the table shows also the values of $\mathfrak{J}_1, \mathfrak{J}_2, \mathfrak{J}_3$ that are calculated equations () and those calculated by Bohr formula (). Again the results of this work for each component and also for each nucleus are much better than the experimental ones.

| A | $\gamma$ | $\beta$ | $\mathfrak{J}_{B1}/\mathfrak{J}_r$ | $\mathfrak{J}_1/\mathfrak{J}_r$ Before | $\mathfrak{J}_1/\mathfrak{J}_r$ After | $\mathfrak{J}_1/\mathfrak{J}_r$ Exp. | $\mathfrak{J}_{B2}/\mathfrak{J}_r$ | $\mathfrak{J}_2/\mathfrak{J}_r$ Before | $\mathfrak{J}_2/\mathfrak{J}_r$ After | $\mathfrak{J}_2/\mathfrak{J}_r$ Exp. | $\mathfrak{J}_{B3}/\mathfrak{J}_r$ | $\mathfrak{J}_3/\mathfrak{J}_r$ Before | $\mathfrak{J}_3$ After | $\mathfrak{J}_3$ Exp. |
|---|---|---|---|---|---|---|---|---|---|---|---|---|---|---|
| 110 | 29 | **0.283** | 0.08 | 0.260 | 0.460 | 0.624 | 0.02 | 0.072 | 0.084 | 0.151 | 0.059 | 0.053 | 0.210 | 0.248 |
| 150 | 10.4 | **0.283** | 0.07 | 0.197 | 0.349 | 0.388 | 0.04 | 0.131 | 0.200 | 0.240 | 0.008 | 0.006 | 0.034 | 0.088 |
| 156 | 7.9 | **0.33** | 0.09 | 0.310 | 0.470 | 0.713 | 0.06 | 0.228 | 0.308 | 0.245 | 0.007 | 0.006 | 0.029 | 0.076 |
| 166 | 9.2 | **0.346** | 0.10 | 0.374 | 0.540 | 0.498 | 0.06 | 0.261 | 0.330 | 0.344 | 0.010 | 0.009 | 0.043 | 0.101 |
| 168 | 8.4 | **0.345** | 0.10 | 0.364 | 0.528 | 0.490 | 0.06 | 0.262 | 0.337 | 0.342 | 0.008 | 0.008 | 0.036 | 0.095 |
| 172 | 4.9 | **0.331** | 0.09 | 0.293 | 0.443 | 0.412 | 0.06 | 0.243 | 0.341 | 0.379 | 0.003 | 0.002 | 0.011 | 0.051 |
| 182 | 10 | **0.241** | 0.05 | 0.118 | 0.245 | 0.366 | 0.03 | 0.080 | 0.144 | 0.227 | 0.005 | 0.003 | 0.022 | 0.054 |
| 184 | 11.3 | **0.234** | 0.05 | 0.111 | 0.237 | 0.310 | 0.02 | 0.071 | 0.129 | 0.207 | 0.006 | 0.004 | 0.026 | 0.074 |
| 186 | 20.4 | **0.207** | 0.04 | 0.090 | 0.217 | 0.351 | 0.01 | 0.039 | 0.070 | 0.131 | 0.015 | 0.009 | 0.062 | 0.081 |
| 188 | 19.9 | **0.193** | 0.04 | 0.073 | 0.188 | 0.280 | 0.01 | 0.032 | 0.062 | 0.120 | 0.012 | 0.007 | 0.052 | 0.100 |
| 190 | 22.1 | **0.184** | 0.03 | 0.065 | 0.178 | 0.256 | 0.01 | 0.026 | 0.051 | 0.090 | 0.016 | 0.008 | 0.057 | 0.113 |